\documentclass[twocolumn,aps,superscriptaddress,showpacs,nofootinbib]{revtex4}
\usepackage{hyperref}
\usepackage{footnote}
\usepackage{amssymb}
\usepackage{amsmath}
\usepackage{graphicx}
\usepackage[normalem]{ulem}
\usepackage{url}
\usepackage{csquotes}
\usepackage{color}

\begin{document}
\title{Binding energy shifts from heavy-ion experiments in a nuclear statistical equilibrium model}

\author{S. Mallik}
\email{swagato@vecc.gov.in}
\affiliation{Physics Group, Variable Energy Cyclotron Centre, 1/AF Bidhan Nagar, Kolkata 700064, India}

\author{H. Pais}
\email{hpais@uc.pt}
\affiliation{CFisUC, Department of Physics, University of Coimbra, 3004-516 Coimbra, Portugal}
\author{F. Gulminelli}
\email{gulminelli@lpccaen.in2p3.fr}
\affiliation{Normandie Univ., ENSICAEN, UNICAEN, CNRS/IN2P3, LPC Caen, F-14000 Caen, France}
\begin{abstract}

Chemical constants extracted from $^{124}$Xe+ $^{124}$Sn collisions at 32 AMeV  are compared to the predictions of an extended Nuclear Statistical Equilibrium model including mean-field interactions and in-medium binding energy shifts for the light ($Z\leq 2$) clusters. The ion species and density dependence of the in-medium modification is directly extracted from the experimental data. We show that the shift increases with the mass of the  cluster and the density of the medium, and we provide a simple linear fit for future use in astrophysical simulations in the framework of the CompOSE data base.
The resulting mass fractions are computed in representative thermodynamic conditions relevant for supernova and neutron star mergers.
A comparison to the results of a similar analysis of the same data performed in the framework of a relativistic mean-field model shows a good agreement at low density, but significant discrepancies close to the Mott dissolution of clusters in the dense medium.
\end{abstract}

\maketitle
 \section{Introduction} \label{sec:intro}

Light clusters such as Hydrogen and Helium isotopes are expected to be abundantly produced in the thermodynamic conditions corresponding to supernova matter and neutron star mergers \cite{Arcones2008,Furusawa2017,Eos_table}, and their abundancy is needed to correctly evaluate the transport properties of hot and dense matter.
However, light clusters properties are modified in the matter. In the case of heavy clusters, that can be treated in the Thomas-Fermi or Hartree-Fock approximation, it is well known \cite{Ravenhall1983,LS} that the excluded volume mechanisms exhaust the most important part of the in-medium effects, and residual modifications can be implemented in the surface tension. However, surface and bulk are not relevant concepts for Hydrogen and Helium isotopes, and more sophisticated few-body calculations should be employed. Such in-medium effects have been computed from a quantum-statistical approach \cite{Roepke2009,Roepke2015,Roepke2021}, but the calculations are limited to a relatively small domain of temperatures and densities. From a phenomenological point of view, a possible solution is the introduction of a density-dependent binding energy shift, which
should be fixed with the help of experimental constraints \cite{HempelPRC91,PaisPRC97} from heavy ion collisions, where such clusters are copiously produced, though in a transient configuration.

A useful observable to pin down the in-medium effects is given the measurement of chemical constants \cite{QinPRL108}. Recently, chemical constants were extracted from INDRA data with a model independent analysis using Bayesian inference \cite{Pais2020}. In \cite{Pais2020}, the extraction of temperature (T) and baryonic density ($n_B$) from the heavy-ion collisions data was performed self-consistently, allowing for arbitrary deviations from the ideal gas limit under the constraint of a common volume for the different particle species.

    A fit of those data was performed within different versions of the relativistic mean field model \cite{Pais2020,Pais2020b}. Interestingly, once the in-medium coupling were fixed to reproduce the experimental data, the different models were shown to produce compatible predictions at relatively low density, but some residual model dependence was observed concerning the location of the Mott dissolution density of the clusters in the dense medium. In this paper, to further study the model dependence of the cluster yields, we extend the study of Ref.~\cite{Pais2020b} to include a model of a different family, that is the extended Nuclear Statistical Equilibrium (NSE) model of Ref.~\cite{Mallik_NuclAstro1}.

{ The paper is organized in the following way:} the model is presented in Section \ref{sec:NSE}, { and} the binding energy shifts are extracted from the experimental data in Section \ref{sec:Results}, where we also give a first comparison of the predicted mass fractions of the clusters, extrapolated to thermodynamic conditions relevant for supernova dynamics, with two different relativistic and non relativistic functionals.
Finally, conclusions are drawn in Section \ref{sec:Conclusions}.

 \section{Nuclear statistical equilibrium model} \label{sec:NSE}
In the extended NSE model \cite{Mallik_NuclAstro1,Raduta}  the densities of clusters as well as free baryons at baryonic number density $n_B=n_n+n_p$, proton fraction $y_p=n_p/n_B$ and temperature $T$ in the grand-canonical formalism are  given by \footnote{we use $k_B=1$ all over the paper.}:
\begin{equation}
n_{q}=\frac{1}{2\pi^2} \Big{(}\frac{2 m^*_{g,q}T}{\hbar^2}\Big{)}^{3/2}F_{1/2}(\eta_{g,q}) + \sum_{A,Z} N_q n_{AZ} , \label{eq:NSE_np}
\end{equation}
where the index $q=n,p$ refers to neutrons and protons. The first term on the right hand side of eq. (\ref{eq:NSE_np})  represents the free proton (neutron) number density $n_{g,q}$  with effective mass   $m^*_{g,q}$  and effective chemical potential $\eta_{g,q}= \left (\mu_{q}-U_{g,q}\right )/T$, 
subject to the density dependent self-consistent mean field $U_{g,q}$ ; the second term of (\ref{eq:NSE_np}) includes the contribution from all bound clusters. The number density of clusters having $N_p=Z$ protons and $N_n=N$  neutrons ($A=Z+N$) can be expressed as
%
\begin{equation}
n_{A,Z}=  (1-u_c)\frac{2J_{AZ}+1}{\lambda^{3}} \exp  - \frac{ 
F_{A,Z}-\sum_{q}\mu_q N_q}{T} , \label{eq:nNZ}
\end{equation}
%
where $\lambda=\hbar[2\pi/(MT)]^{1/2}$ is the de Broglie wavelength, with $M=N m_n + Z m_p$ the bare bound ion mass,  $J_{AZ}$ the ground state spin, and  the term $(1-u_c)$ is an excluded volume correction that modifies the space integration associated to the center of mass free energy of each cluster with $u_c=\frac{V_{c}^{tot}}{V^{tot}}$ ($V_{c}^{tot}$ is the total volume occupied by all bound clusters and $V^{tot}$ is the total volume). For the internal Helmholtz free energy $F_{A,Z}$  of Hydrogen and Helium isotopes, no excited states are considered
and we have:
\begin{equation}
F_{A,Z}=E_{exp} + \Delta E(n_B,T,A,Z) , \label{eq:binding}
\end{equation}
where $E_{exp}$ is the measured ground state energy, and $ \Delta E$ is the binding energy shift that we want to extract from the comparison with the experimental chemical constants.
For heavier clusters ($Z>2$) we suppose that the excluded volume mechanism  exhausts the in-medium effects, and the internal free energy  is decomposed as,
\begin{equation}
F_{A,Z}= F_{A,Z}^{bulk}+ F_{A,Z}^{surf}+F_{A,Z}^{coul} , \label{eq:Helmholtz}
\end{equation}
where $F_{A,Z}^{bulk}$ is the bulk contribution, corresponding to a volume $V_{c}=A/n_c$ of nuclear matter at density $n_c=n_{c,n}+n_{c,p}$ ($n_{c,n}$ and  $n_{c,p}$ are the neutron and proton density inside the bound cluster) and  isospin asymmetry  $\delta_c= (n_{c,n}-n_{c,p})/n_c$. The evaluation of the cluster free energy requires dealing with the well known problem of double
counting of continuum states in the system partition sum \cite{Tubbs}. To this aim, a continuum subtraction procedure was introduced in \cite{Mallik_NuclAstro1}. The analytical form of continuum subtracted  bulk part of Helmholtz free energy is given by,
%
\begin{eqnarray}
 F_{A,Z}^{bulk}&=&V_{c} \bigg{ [}v(n_c,\delta_c)-v(n_g,\delta_g) \bigg {]} \nonumber \\
&-& V_{c}  \sum_{q=n,p} \left ( U_{c,q}n_{c,q}-U_{g,q}n_{g,q}\right ) \nonumber \\
&-&\frac{2 V_c}{3}\sum_{q=n,p}\bigg{[}\xi_{c,q}-\xi_{g,q}+\mu_{q}N_q \bigg{]},
\end{eqnarray}
%
where 
$\xi_{c,q}=\frac{3h^2}{2\pi m^*_{c,q}}\bigg{(}\frac{2\pi m^*_{c,q}T}{h^2}\bigg{)}^{5/2}F_{3/2}(\eta_{c,q})$ is the kinetic energy density of the cluster with the baryon of type $q=n,p$ in uniform matter at density $n_{c}$ and asymmetry $\delta_{c}$ . $v(n_c,\delta_c)$ and $v(n_g,\delta_g)$ are the potential energy density of the bound cluster and of the free nucleon gas, respectively. The bulk part of the cluster functional, and the corresponding mean-field are determined from meta-modelling of the EoS \cite{Margueron2018a} with Sly5 parameters \cite{Chabanat}. The details can be found in Appendix A of Ref.~\cite{Mallik_NuclAstro1}. The finite size effect of the cluster is introduced by the surface part of Helmholtz free energy given by,
\begin{eqnarray}
 F_{A,Z}^{surf}&=&4\pi r^2_c A_c^{2/3}\sigma(y_{c,p},T) ,
\label{Surface}
\end{eqnarray}
where $A_c=A+(\rho_{g,p}+\rho_{g,n}) V_c$,  $r_c={\{} 3/(4\pi n_c){\}}^{1/3}$, $y_{c,p}=Z/A$ and $\sigma(y_{c,p},T)$ is the surface tension at finite temperature $T$ and proton fraction $y_{c,p}$ \cite{Mallik_NuclAstro1,LS,Carreau2019}. The third term of Eq.~(\ref{eq:Helmholtz}) represents the temperature independent Coulomb energy given by
\begin{equation}
 F_{A,Z}^{coul}=\frac{3}{5}\frac{e^2Z^2}{4\pi\epsilon_0}(1-f_{WS})\left (\frac{4\pi}{3V_c}\right )^{1/3} ,
\end{equation}
where $f_{WS}$ is the Wigner-Seitz approximation factor introduced to treat long range Coulomb interaction in the statistical ensemble formalism \cite{Francesca2015}.
\begin{figure}[!hbtp]
\begin{center}
\includegraphics[width=0.44\textwidth]{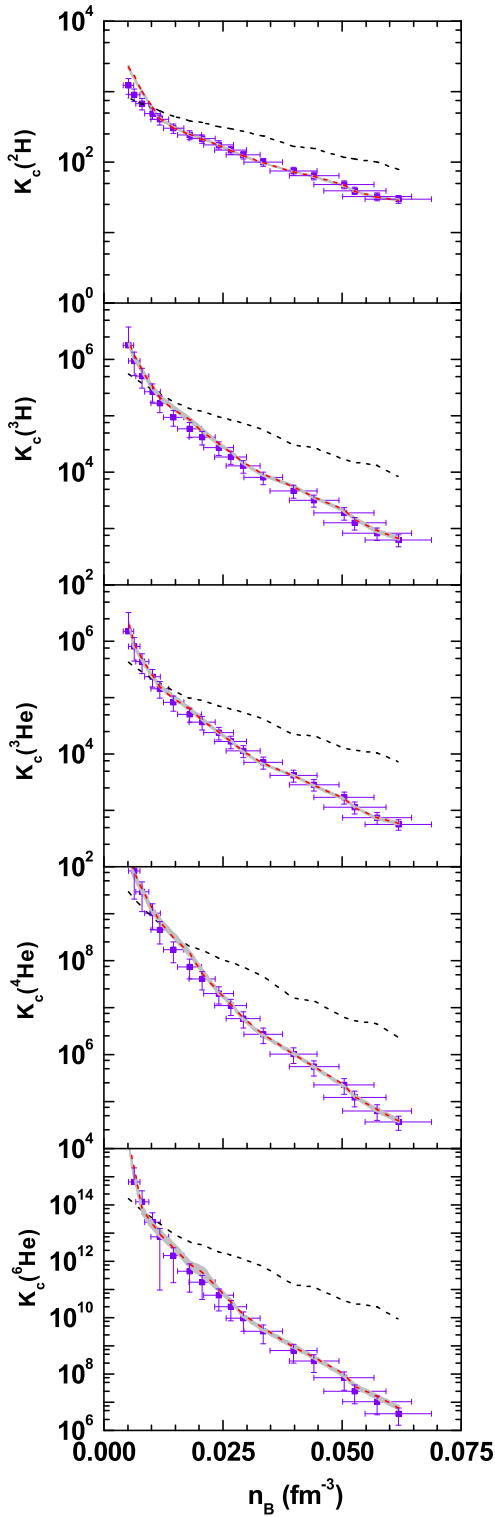}
\caption{Density dependence of chemical equilibrium constants calculated from the NSE model without in-medium correction (black dashed lines), and  with an in-medium correction optimized on the data. Mean values are represented by red dashed lines, whereas the shaded region represents the $2\sigma$ uncertainty interval. The violet squares represent the experimental values in the $^{124}$Xe on $^{124}$Sn reaction at 32 AMeV from Ref.~\cite{Pais2020}.}
\label{Chemical_constant}
\end{center}
\end{figure}

 \section{Results} \label{sec:Results}

%
%

In a well-defined thermodynamic equilibrium condition characterized by the temperature $T$, total baryonic density $n_B$ and proton fraction $y_p$ , the equilibrium chemical constant $K_c(A,Z)$ of a cluster of mass number $A$ and atomic number $Z$ can be defined as
\begin{equation}
K_c(A,Z)=\frac{n_{AZ}}{n_{g,p}^Z n_{g,n}^{A-Z}} ,
\end{equation}
where $n_{AZ}$, $n_{g,p}$ and $n_{g,n}$ are the densities of the specific cluster of mass $A$ and charge $Z$, free protons and free neutrons respectively. The  thermodynamic variables $(T,n_B,y_p)$ are extracted from the experimentally measured multiplicities for different surface velocity bins in $^{124}$Xe  on $^{124}$Sn central collision reactions at 32 AMeV performed by INDRA collaboration \cite{Bougault2020,Pais2020}, and used as input for the NSE model calculation. The baryonic density dependence of chemical equilibrium constant  for $^{2}$H, $^{3}$H, $^{3}$He, $^{4}$He and  $^{6}$He is initially investigated from the NSE model without in-medium correction, that is $\Delta E=0$ in Eq.~(\ref{eq:binding}), and the result is noted $K_c^{free}$. The comparison to
the experimental data $K_c^{expt}$ is shown in Fig.~\ref{Chemical_constant}. The NSE calculation 
nicely reproduces the data at low density, but it increasingly overestimates the cluster population as the density increases. This confirms the need of including in-medium corrections to the experimental binding energies of the light clusters, as already observed in Ref.~\cite{HempelPRC91}.

\begin{figure}[!hbtp]
\begin{center}
\includegraphics[width=0.45\textwidth]{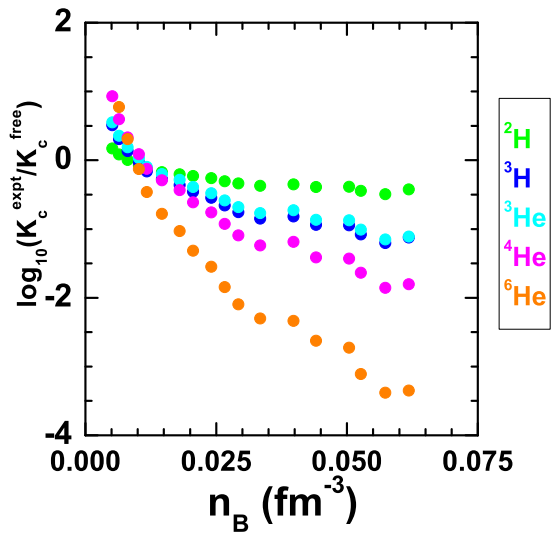}
\caption{Density dependence of the experimental deviation from the vacuum energy assumption $log_{10}\frac{K_c^{expt}}{K_c^{free}}$ for the different light clusters.}
\label{Chemical_constant_expt_to_theo_ratio}
\end{center}
\end{figure}

To better visualize the deviation, Fig.~\ref{Chemical_constant_expt_to_theo_ratio} represents the density dependence of the ratio $\log_{10}\frac{K_c^{expt}}{K_c^{free}}(A,Z)$ for the above mentioned light clusters.  We can observe   that  this ratio is almost the same for the isobars $^{3}$H and $^{3}$He, while a clear mass hierarchy is observed.  This suggests  that the in-medium correction factor strongly depends on $A$ but the dependence on the isospin $I=N-Z$ can be neglected.
We therefore propose a simple three-parameters expression for the correction:  $\Delta E=a_1 +a_2 A^{a_3} $, where $a_k(n_B,T)$, $k=1,2,3$,  are fitting parameters which can be optimized to the data at each baryonic density point\footnote{It will be very important to have data on more asymmetric systems to confirm the similar behavior of $^3$H and $^3$He, since the assumption of isospin independence might have important consequences in the extrapolation to the very neutron-rich matter involved in astrophysical sites where those clusters are produced.}.

The likelihood probability at each baryonic density   and temperature  can be defined as:
\begin{eqnarray}
P(\vec a)=\mathcal{N}\exp\bigg{(}-\frac{1}{2}{\sum_{AZ}\Bigl{\{}\frac{\log_{10} K_c^{theo}-\log_{10} K_c^{expt}}{\Delta \log_{10} K_c^{expt}} \Bigr{\}}^2}\bigg{)}.
\label{likelihood_probability}
\end{eqnarray}
Here, $\mathcal{N}$ is the density, proton fraction and temperature dependent normalization constant, $K_c^{theo}(A,Z,\vec a)$, with $\vec a=\{a_1,a_2,a_3\}$,  is the equilibrium chemical constant of a cluster of mass number $A$ and atomic number $Z$ with in-medium correction factor $\Delta E=a_1+a_2A^{a_3}$, and  $\Delta \log_{10} K_c^{expt}$ indicates the  experimental standard deviation.

\begin{figure}[!hbtp]
\begin{center}
\includegraphics[width=0.45\textwidth]{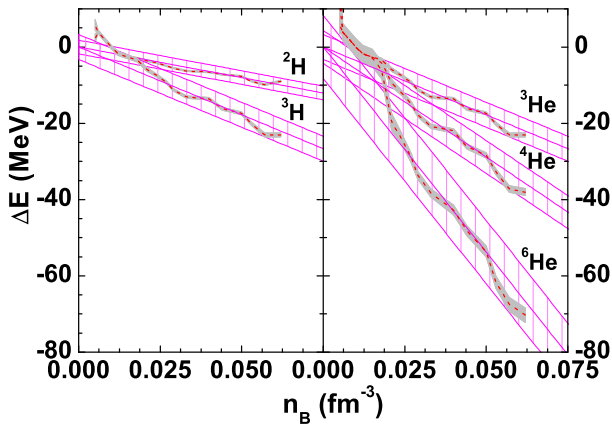}
\caption{Density dependence of the in-medium correction factor of the nuclear binding energy for light clusters extracted from the NSE model optimized on the experimental data.
The width of the thin band represents $2\sigma$ uncertainty intervals. A linear fit of these curves is given by the magenta solid lines. The standard deviation of the fit is given by the magenta shaded region.   }
\label{Binding_correction}
\end{center}
\end{figure}

The expectation value of any physical quantity $X$ can be determined from this probability distribution as:
\begin{eqnarray}
\langle X \rangle=\int \int \int P(\vec a)X(\vec a)d\vec a ,
\label{average_value}
\end{eqnarray}
with standard deviation $\sigma_X=\sqrt{{\langle X^2 \rangle}-\langle X \rangle^2}$.
The resulting in-medium corrected chemical constants are presented by shaded areas in Fig.~\ref{Chemical_constant}, which shows the quality of the fit.

\begin{figure*}[!hbtp]
\begin{center}
\includegraphics[width=.9\textwidth]{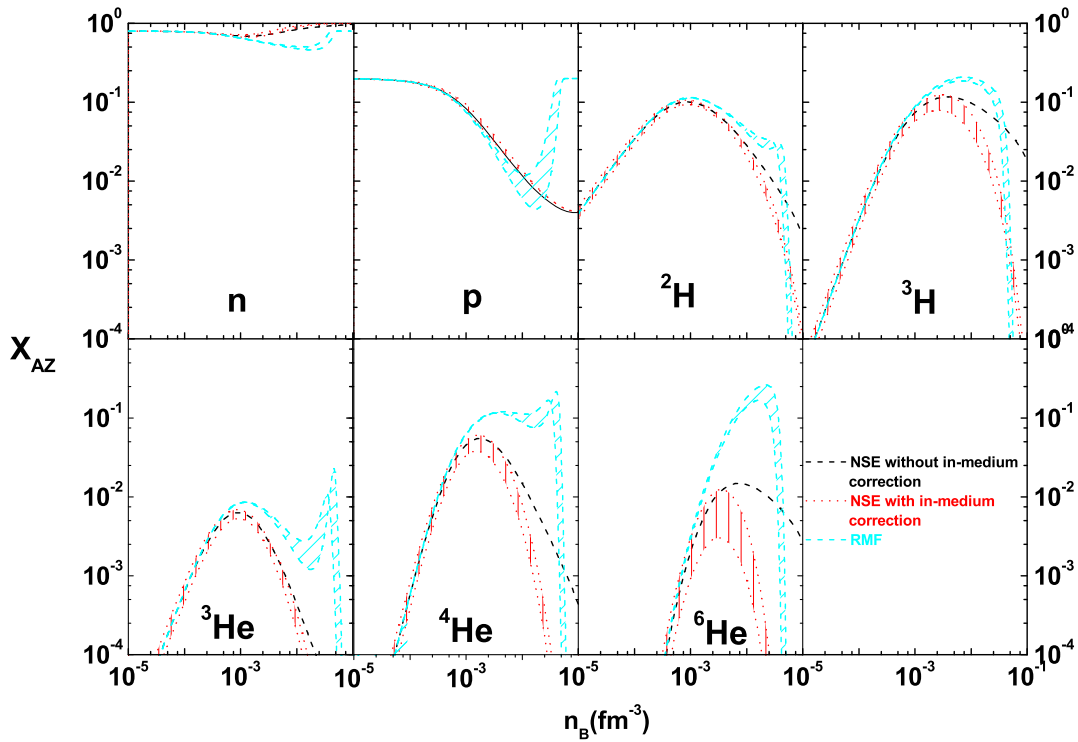}
\caption{Density dependence of the mass fraction of light clusters calculated from different models at a representative constant temperature $T=5$ MeV and proton fraction $y_p=0.2$. See text for details.}
\label{model_comparison}
\end{center}
\end{figure*}

Similarly, the density dependence of the optimized binding energy shifts $\Delta E$ is shown in Fig.~\ref{Binding_correction}.

In this analysis, the binding energy shifts are determined only at the points where experimental chemical constants as well as thermodynamic  parameters $(T,n_B,y_p)$ are known. The corresponding values of temperature, density and proton fraction are displayed in Table \ref{tab:tab1}.
\begin{table}[htb]
\begin{center}
\begin{tabular}{|c|c|c|}
\hline
 $n_B$ (fm$^{-3}$) & $T$ (MeV) &  $y_p$  \\
\hline
5.14$\times 10^{-3}$	&	5.12 	&	  0.43 	  	\\
6.40$\times 10^{-3}$	&	5.26 	&	  0.42 	  	\\
8.05$\times 10^{-3}$	&	5.44 	&	  0.42 	  	\\
1.02$\times 10^{-2}$ 	&	5.67 	&	  0.42	  	\\
1.17$\times 10^{-2}$	&	5.82 	&	  0.42 	  	\\
1.46$\times 10^{-2}$	&	6.13 	&	  0.42 	  	\\
1.80$\times 10^{-2}$	&	6.39 	&	  0.42 	  	\\
2.06$\times 10^{-2}$	&	6.48 	&	  0.43 	  	\\
2.41$\times 10^{-2}$	&	6.69 	&	  0.43	  	\\
2.66$\times 10^{-2}$	&	6.79 	&	  0.43	  	\\
2.92$\times 10^{-2}$	&	6.92 	&	  0.43		\\
3.34$\times 10^{-2}$    &    7.17       &      0.43		\\
3.98$\times 10^{-2}$    &    7.89       &      0.43		\\
4.41$\times 10^{-2}$    &    7.91       &      0.44          \\
5.04$\times 10^{-2}$    &    8.48       &      0.45		\\
5.27$\times 10^{-2}$    &    8.56       &	  0.45		\\
5.73$\times 10^{-2}$    &    8.59       &      0.46		\\
6.18$\times 10^{-2}$    &    9.06       &      0.48	\\	
\hline
\end{tabular}
\end{center}
\caption{\label{tab:tab1} Correlation between the average density, temperature and proton fraction as extracted from the experimental analysis of Ref.~\cite{Pais2020a}.
}
\end{table}

Qualitatively, we expect the effective binding energies to decrease with density, while thermal effects should lead to higher clusters center of mass momentum,  reduced Pauli blocking effects, and therefore a smaller suppression \cite{Roepke2009}. Since the experimental suppression increases with the simultaneous increase of $n_B$ and $T$,  we deduce that the suppression effect of the density seems to dominate over the thermal effects, at least in the temperature interval explored in the experiment.
This is certainly a crude approximation, and we hope that a combined analysis of different data sets in the next future, corresponding to different $T(n_B)$ trajectories of the expanding system, will allow us to disentangle the temperature, density, and proton fraction dependence.
In order to determine $\Delta E$  over a wide and continuous density region for further applications within the CompOSE\cite{compose} database, we perform a linear fit  $\Delta E=m \cdot n_B$. The results are also shown in Fig.~\ref{Binding_correction}. The mean values and standard deviation of the slope parameter $m$ are reported in Table 2.

\begin{table}[!b]
\begin{center}
\begin{tabular}{|c|c|c|}
\hline
 & \multicolumn{2}{c|}{$\Delta E/n_B$ (MeVfm$^3$)}  \\
\cline{2-3}
Cluster & Mean & Standard  \\
 & & Deviation  \\
\hline
$^{2}$H &-160.5&23.3\\
$^{3}$H & -355.5& 42.7\\
$^{3}$He & -355.5& 42.7\\
$^{4}$He & -576.8& 56.3\\
$^{6}$He & -1075.8& 108.4\\
\hline
\end{tabular}
\end{center}
\caption{Mean value and standard deviation of the binding energy shift $m=\Delta E/n_B$  in the linear approximation  for $^{2}$H, $^{3}$H,  $^{3}$He,  $^{4}$He,  $^{6}$He isotopes.}
\label{tab:tab2}
\end{table}

It is interesting to remark that a qualitatively similar density dependence for the binding energy  shifts was obtained in Ref.\cite{PaisPRC97}. However, a direct comparison of the $\Delta E$ corrections of Fig.~\ref{Binding_correction} with the effective masses of Ref.\cite{PaisPRC97} is not possible because of the different mechanisms of cluster suppression in the two formalisms. The compatibility of the different approaches can only be established by comparing the cluster abundances, as we will do below.

This parameterised form of the in-medium binding energy shifts obtained from the optimization of the NSE model to the experimental chemical constants, can now be used for astrophysical predictions.  As an example, we have studied the effect of including  the shifts $\Delta E$ on  mass fractions in warm stellar matter.
The NSE calculation is performed for two cases: (i) without in-medium correction $\Delta E=0$; (ii) supposing a linear density dependence of $\Delta E$, using the parameters given in Table 2.

A representative calculation  is performed at constant temperature $T=5$ MeV and proton fraction $y_p=0.2$,  and varying the baryonic density $n_B$  from $10^{-5}$ fm$^{-3}$ to $0.1$ fm$^{-3}$. The mass fraction of various clusters are defined as:
\begin{equation}
X_{AZ}=\frac{A n_{AZ}}{\sum_i A_i n_{A_iZ_i} }, \label{eq:fraction}
\end{equation}

where the sum in the denominator is performed over $n$, $^{1,2,3}H$, $^{3,4,6}He$. The results are  shown in Fig.~\ref{model_comparison}.
The large error bars are due to the uncertainty in the linear fitting of the binding energy shift. We expect that new data in different thermodynamic conditions in the near future will allow discriminating between the temperature and density dependence of the correction, and better pin down the density dependence.

We can see that the introduction of the binding energy shifts leads to an important reduction of the mass fractions of the clusters, which are dissolved in dense matter at a density varying between $\approx 2\times 10^{-2}$ and $5\times 10^{-2}$ fm$^{-3}$, depending on the isotope. In spite of the highest binding of the $\alpha$ particle, the highest Mott density is observed for $^3$H, which is also the most abundant bound cluster in this neutron-rich matter. Since the in-medium effects are seen to approximately scale with the  ion mass number, the deuteron energy shift is the least important one.
As a consequence, this loosely bound cluster is the second most abundant one after the triton.

Abundancies of light clusters in warm stellar matter were recently calculated
from this same NSE model in Ref.\cite{Mallik_NuclAstro1}. It was shown that at moderate temperatures and low proton fractions, resonant states beyond the neutron dripline such as $^{4,5,6,7}$H and $^{7,8,9,10}$He could be dominant in the equilibrium, if vacuum energies are considered for all the light species. We expect that in-medium effects should suppress the resonant population at least as much as the stable particles. In the absence of any experimental or theoretical constraint on the mass shifts of unstable states, for the calculation displayed in Fig.~\ref{model_comparison} we have suppressed the isotopes beyond the driplines from the partition sum. Including them with vacuum energies only marginally changes the results of Fig.~\ref{model_comparison}, because of the chosen normalization of the mass fractions Eq.(\ref{eq:fraction}). We also stress that the inclusion of exotic species is not influent at all in the prediction of the chemical constants Fig.\ref{Chemical_constant}.

The calculation of the mass fractions also allows assessing the model dependence of the theoretical predictions. The same INDRA data shown in Fig.~\ref{Chemical_constant}
were used in Refs.~\cite{Pais2020,Pais2020b} to calibrate the in-medium couplings of the nucleons bound in clusters to the effective mesons that are responsible for the strong interaction in the Relativistic Mean Field (RMF) formalism. Specifically, a  linear correction for the scalar coupling acting on nucleons bound in a cluster of mass $A$ was introduced, $g_s(A)=x_s A g_s$, with $g_s$ the scalar coupling of nucleons in homogeneous matter, and $x_s$ a free parameter. For a fixed RMF functional, it was shown in Ref.~\cite{Pais2020a} that a unique value of $x_s$ allowed to give a reasonable reproduction of the whole density dependence of the chemical constants. Moreover, different RMF functionals were shown in Ref.~\cite{Pais2020b} to produce compatible predictions up to a density $n_B\approx 0.05$ fm$^{-3}$, once an optimized value  for the parameter $x_s$ is adopted for each of them. However, some model dependence was observed at higher density, particularly concerning the value of the dissolution density of the clusters in the dense medium.

The mass fractions of the different cluster species, as computed from the DDME2 \cite{DDME2} RMF functional with the optimized $x_s$ value,  $x_s=0.93\pm0.02$, are calculated in the same thermodynamic conditions as for the NSE model, and the results are shown in Fig.~\ref{model_comparison}. The optimization on the experimental data was done in Ref.\cite{Pais2020b}, following the protocol introduced in Ref.\cite{PaisPRC97}.

Similar to Ref.~\cite{Pais2020b}, we can observe a very nice agreement at low density, but considerable deviations appear close to the dissolution density.

Different reasons can be invoked to explain those deviations. First, we cannot exclude an effect of the equation of state, and different functionals within this same NSE treatment should be considered.
Moreover, not only the mechanism of cluster suppression is very different in the non-relativistic NSE with respect to the RMF, but also the degrees of freedom are not the same: only homogeneous matter and $Z\leq 2$ clusters are considered in the RMF calculation, while clusters of all sizes up to $Z=100$ are included in the NSE. It has been shown  in previous works
that the introduction of heavier $5\le A \le 12$ clusters \cite{Pais2019} only marginally modifies the chemical constants, and that the influence of heavy pasta-like clusters is also a minor correction\cite{Avancini2017}, however a more detailed comparison is needed to settle this point.\\
\indent
Finally, the corrections extracted at the limited set of points in the $(n_B,T,y_p)$ space accessible to the experiment are extrapolated to very different thermodynamic conditions in Fig.~\ref{model_comparison}. This could indicate that more complex temperature and isospin effects should be considered.\\
\indent
Further theoretical studies, more extensive comparisons with the experimental data, and extra constraints at higher density will be needed to clarify the model dependence of the results.

\section{Conclusions}\label{sec:Conclusions}

In this paper, we have determined the binding energy shifts of light clusters in a dense thermalized medium, by a comparison of the extended NSE model of Ref.~\cite{Mallik_NuclAstro1} using the Sly5 functional, to the chemical constants measured by the INDRA collaboration in $^{124}$Xe$+^{124}$Sn collisions at 32 AMeV.
Density and particle number dependent in-medium binding energy shifts are explicitly included in the partition sum, and their value for each  ion species and density point is determined by imposing that the predicted chemical constants coincide with the measured ones within error bars.
In qualitative agreement with previous works \cite{Pais2020,Pais2020b}, we find that the shifts increase with increasing mass of the  clusters and increasing density of the surrounding medium, while no isospin dependence is observed.

The density dependence of the shifts is fitted by a simple parametrization that can be easily incorporated in future simulations of dense matter in astrophysical conditions. In particular, they will be used to produce complete tables of general purpose equations of state for the CompOSE data base. These calculations are in progress.

The predictions of the model were further compared to the ones obtained with the RMF formalism using the DDME2 functional, and in-medium scalar meson couplings optimized to the same data set.
A first representative calculation of mass fractions at $T=5$ MeV and proton fraction $y_p$=0.2 shows good agreement at low density, but considerable deviations close to the dissolution density of clusters in the dense medium.

\section{Acknowledgements}
We thank R.
Bougault and D. Gruyer of LPC Caen and C. Provid\^encia of  University of Coimbra for valuable discussions. This work was partly supported by the FCT (Portugal) Projects No. UID/FIS/04564/2020, and POCI-01-0145-FEDER- 029912, by PHAROS COST Action CA16214, and by “IFCPAR/CEFIPRA” Project No. 5804-3.
H.P. acknowledges the grant CEECIND/03092/2017 (FCT, Portugal).

\end{document}